\documentstyle[prb,twocolumn,epsf,aps]{revtex}
\tighten
\begin{document}

\title{Coherent patterns and self-induced diffraction of electrons on
a thin nonlinear layer}
\author{O. M. Bulashenko,$^{1,2}$\cite{byline} V. A. Kochelap,$^2$
and L. L. Bonilla$^1$}
\address{
$^1$Universidad Carlos III de Madrid, Escuela Polit\'ecnica Superior,
Butarque 15, E-28911 Legan\'es, Spain \\
$^2$Department of Theoretical Physics, Institute of Semiconductor
Physics, National Academy of Sciences, Kiev 252028, Ukraine}
\address{\rm (1 February 1996)}
\address{\mbox{ }}
\address{\parbox{14cm}{\rm \mbox{ }\mbox{ }\mbox{ }
Electron scattering on a thin layer where the potential depends
self-consistently on the wave function has been studied.
When the amplitude of the incident wave exceeds a certain threshold,
a soliton-shaped brightening (darkening) appears on the layer causing
diffraction of the wave.
Thus the spontaneously formed transverse pattern can be viewed
as a self-induced nonlinear quantum screen.
Attractive or repulsive nonlinearities result in different phase
shifts of the wave function on the screen, which give rise to quite
different diffraction patterns.
Among others, the nonlinearity can cause self-focusing of the incident
wave into a ``beam'', splitting in two ``beams'', single or double
traces with suppressed reflection or transmission, etc.
}}
\address{\mbox{ }}

\maketitle

\makeatletter
\global\@specialpagefalse
\def\@oddhead{\underline{Physical Review B: July 15, 1996\hspace{306pt}
REV\TeX\mbox{ }3.0}}
\let\@evenhead\@oddhead
\makeatother

\narrowtext

The spontaneous formation of spatial structures (patterns) due to
nonlinearity is well known for dissipative systems driven away from
equilibrium.  \cite{cross93}
In solid state physics those patterns have been mostly studied in the
regime governed by classical macroscopic processes, \cite{ICPF95}
where quantum coherence effects were not important.
In this paper we predict the spontaneous formation of {\em quantum
coherent non-dissipative patterns} in semiconductor heterostructures
with nonlinear properties.

Since the Schr\"odinger equation is linear, the nonlinearity appears
in quantum systems due to the many-body effects and/or the coupling
with the environment.
In a mean-field approximation this problem can be traced to the
self-consistent Schr\"odinger equation with the Hamiltonian
$H = -\frac{\hbar^2}{2m} \nabla^2 + V({\bf r}) + V_{\rm eff}
[|\psi ({\bf r})|^2]$, where in addition to the external potential
$V({\bf r})$ the self-consistent potential $V_{\rm eff}$ is
introduced, representing a nonlinear response of the medium.
\cite{callaway}
The potential $V_{\rm eff}$ depends on the probability $|\psi ({\bf
r})|^2$ of the carrier to be located at ${\bf r}$.
When (in a weakly-nonlinear case) it is proportional to that
probability, the resultant equation for a single-particle
wave-function $\psi ({\bf r})$ is the so-called nonlinear
Schr\"odinger equation (NSE) with a cubic term \cite{makhankov84}
encountered in different contexts of the solid state physics:
(i) the polaron problem, \cite{pekar} where the strong electron-phonon
interaction deformates the lattice thereby providing an attractive
potential; \cite{smondyrev95}
(ii) the magnetopolaron problem \cite{furdyna88} in semimagnetic
semiconductors, where the exchange interaction between the carrier
spin and the magnetic impurities leads also to an effective attractive
potential; \cite{hawrylak91,benoit95}
(iii) Hartree-type interaction between electrons, giving a repulsive
potential, \cite{jona-lasinio} and others. \cite{makhankov84}

Motivated by the great progress in heterostructure fabrication, some
important results have been obtained recently in the framework of the
cubic NSE for the situations when the nonlinearities are concentrated
in thin semiconductor layers modeled by $\delta$-potentials.
\cite{hawrylak91,malysheva92,sun95,malomed93}
Among these results, we may mention the multiplicity of stable states found
in different physical situations for which tunneling is important: 
an array of semimagnetic quantum dots, \cite{hawrylak91}
a quantum molecular wire, \cite{malysheva92} a doped superlattice
formed by $\delta$-barriers.  \cite{sun95}
Another is the oscillatory instability of the flux transmitted through
the nonlinear layer. \cite{malomed93}
It should be noted, however, that all these results are restricted to
one-dimensional spatial supports, which means that the longitudinal
and transverse degrees of motion are assumed to be decoupled.
Disregarding that assumption in this paper, we show that considering
additional spatial dimensions opens up the possibility of
qualitatively new nonlinear phenomena -- spontaneous formation of
spatial transverse patterns, which are quantum-mechanically coherent.

Consider a thin layer in the $xy$-plane with the concentrated
nonlinearity.  We model the layer by using the $\delta$-function which
simplifies greatly the calculations without modifying qualitatively
the results.
Keeping in mind possible pattern formation and analogy with the
optics, the layer can be thought of as a screen.
The steady-state scattering problem for the thin $\delta$-layer is
governed by the NSE
\begin{eqnarray} \label{schr-eq}
-\frac{\hbar^2}{2m} \Delta \psi ({\bf r}) + (A + B |\psi ({\bf r})|^2)
\; \delta (z) \; \psi ({\bf r}) =E \psi ({\bf r}).
\end{eqnarray}
The external potential $A$ is allowed to be of the both signs, i.e.,
$A>0$ if it is a barrier and $A<0$ if it is a well.
$B$ is the strength of the nonlinear potential: $B<0$ for the
attractive and $B>0$ for the repulsive interaction.
We do not specify the concrete physical model, because our results
could be applicable to any of the above mentioned systems, although
the most feasible candidates for the attractive case are believed to
be semimagnetic heterostructures like CdTe/CdMnTe and CdTe/HgCdMnTe,
where both, the height $A$ of the barrier and the strength $B$ can be
varied by choosing the alloy composition.
In addition, $A$ can be tuned by an external magnetic field.
The expressions for $A$ and $B$ can be found elsewhere.
\cite{hawrylak91}
The repulsive case is an idealization of the situation considered in
Ref.\ \onlinecite{jona-lasinio}; see Ref.\ \onlinecite{sun95}.

We seek the solution in the form

\begin{equation} \label{scat-sol}
\psi (x,y,z)= \left\{ \begin{array}{lr}
a e^{ikz} + b(x,y,z) \; e^{-ikz}, & z < 0 \\
c(x,y,z) \; e^{ikz}, & z > 0
\end{array} \right. \end{equation}
where the amplitude $a$ of the incident wave is fixed (real), and the
electron energy $E=\frac{\hbar^2 k^2}{2m}$.
We assume that there is no current inflow along the screen (the only
inflow into the system is from $z=-\infty$). Thus only those solutions
satisfying the condition of zero inflow at $z$=0, $x,y\to\pm\infty$
will be considered.

\begin{figure}
\vspace*{0.2cm}
\setlength{\epsfxsize}{8cm}
\centerline{\mbox{\epsffile{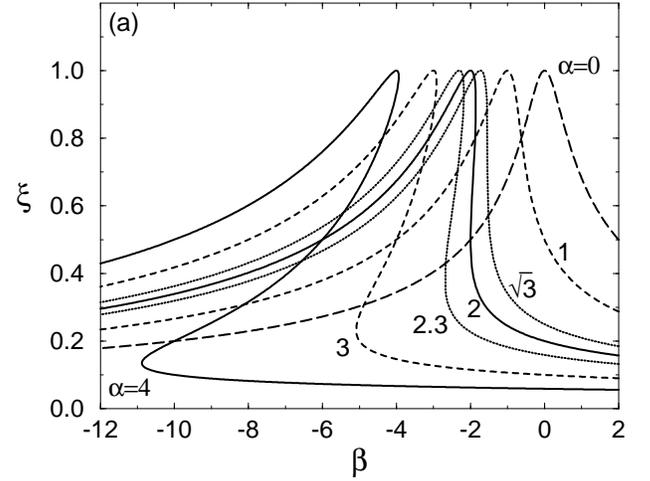}}}
\vspace{0.5cm}
\setlength{\epsfxsize}{8cm}
\centerline{\mbox{\epsffile{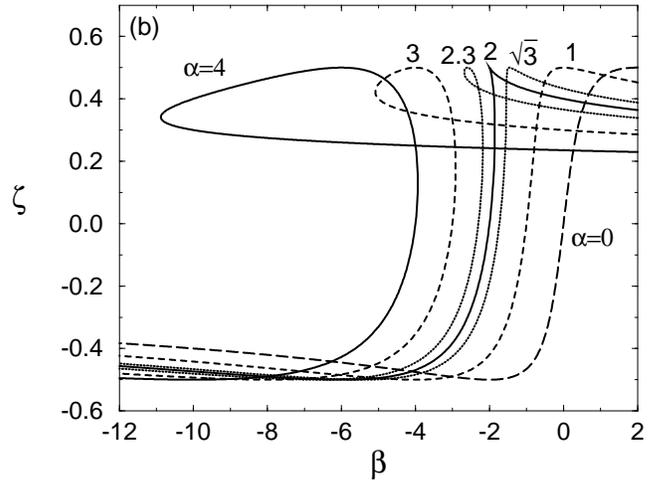}}}
\caption{
Real (a) and imaginary (b) parts of the transmitted amplitude
$\tilde{c}$ as functions of $\beta$ for a uniform solution and
different values of $\alpha$. Only case $\cal A$ is shown.
The case $\cal R$ can be obtained by replacing $\alpha\to - \alpha$,
$\beta\to - \beta$.
}\label{roots}\end{figure}

It is convenient to write (\ref{scat-sol}) and (\ref{schr-eq}) in
dimensionless form by means of the definitions:
$\tilde{x}=\sqrt{2}k x, \tilde{y}=\sqrt{2}k y$, $\tilde{z}= k z$,
$\tilde{b}= b/a$, $\tilde{c}=c/a$.
Insertion of (\ref{scat-sol}) into Equation (\ref{schr-eq})
for $z \neq 0$ yields

\begin{equation} \label{half}
\begin{array}{ll}
\Delta_{\perp} \tilde{b} + \frac{1}{2}\,\partial_{\tilde{z}\tilde{z}}
\tilde{b} - i \;\partial_{\tilde{z}} \tilde{b} = 0, & \tilde{z}<0 \\
\Delta_{\perp} \tilde{c} + \frac{1}{2}\,\partial_{\tilde{z}\tilde{z}}
\tilde{c} + i \;\partial_{\tilde{z}} \tilde{c} = 0, & \tilde{z}>0.
\end{array}
\end{equation}
By using the continuity of the wave-function $\psi$, one gets at
$z$=0:

\begin{equation} \label{screen-full}
\partial_{\tilde{z}} \tilde{c} - \partial_{\tilde{z}} \tilde{b} +
2 i\; (\tilde{c}-1) = 2 (\alpha + \beta |\tilde{c}|^2)\;\tilde{c},
\end{equation}
and $\Delta_{\perp} \tilde{b}$=$\Delta_{\perp} \tilde{c}$.
Here $\alpha$=$m A/(\hbar^2 k)$ and $\beta$=$m B a^2/(\hbar^2 k)$.
Eqs.\ (\ref{half}) and (\ref{screen-full}) have spatially uniform
solutions $\tilde{c}$=$\xi + i \zeta$ such that $\zeta =
- \alpha \xi - \beta \xi^2$, $|\tilde{c}|^2 =\xi$, and
\begin{equation} \label{cubic}
\beta^2\xi^3 + 2 \alpha\beta\xi^2 + (\alpha^2+1)\xi - 1 = 0.
\end{equation}
A straightforward analysis of this equation demonstrates that there is
only one real root for $\alpha^2<3$ and there are three real roots
under the conditions:  $\alpha^2>3$, $\alpha\beta<0$, and
$\beta^{-} < \beta< \beta^{+}$ with $\beta^{\mp}=\frac{2}{27}
[ \mp (\alpha^2-3)^{3/2} - \alpha^3 - 9\alpha]$.
Thus multiple solutions are expected for two cases: the barrier
($\alpha>0$) with attractive nonlinearity ($\beta<0$) (case $\cal A$);
the quantum well ($\alpha<0$) with repulsive nonlinearity ($\beta>0$)
(case $\cal R$). Taking $\beta$ as a control parameter these solutions
are depicted in Fig.\ \ref{roots} for different $\alpha$.
Notice that we obtain up to three coexisting uniform solutions for
different values of $\alpha$: $Z$-shaped curves $\xi(\beta)$
(if $\alpha>\sqrt{3}$) and $S$-shaped ($\sqrt{3}<\alpha<2$) or
loop-shaped (if $\alpha>2$) curves $\zeta(\beta)$.
At $\alpha$=2 there is a cusp of the maximum of the $\zeta(\beta)$
curve. The peaks in Fig.\ \ref{roots}(a) correspond to maxima of the
transmission for which $|\tilde{c}|^2$=$\xi$=1 and $\beta$=$-\alpha$.
Since $\beta\propto a^2$, multiple solutions exist on a certain
interval of incident wave amplitudes for any strength of the
nonlinearity $B$.
The threshold values $\alpha$=$mA/k\hbar^2$=$\pm\sqrt 3$ for
multiplicity of uniform solutions can be achieved by varying the
barrier height (well depth) and/or the energy of the incident wave.
Three uniform solutions coalesce at the tricritical parameter values:
$\alpha_0$=$\pm\sqrt{3}$, $\beta_0$=$\mp 8\sqrt{3}/9$, $\xi_0$=3/4,
$\zeta_0$=$\mp\sqrt{3}/4$.
Hereafter we use the upper sign for case $\cal A$ and the lower sign
for case $\cal R$.

We shall perform now a small-amplitude perturbation analysis of Eqs.\
(\ref{half}) and (\ref{screen-full}) near the tricritical point.
As a result we will find simple amplitude equations that will be
solved in two particular cases of interest: (a) $y$-independent
solutions, and (b) axisymmetric solutions.

Let $\alpha = \alpha_0 \pm \delta$, $\beta=\beta_0\mp\gamma$
with $\delta>0$, $\gamma>0$ and $\delta, \gamma\ll 1$.
We look for small nonuniform solutions: $\tilde c = \xi + i \zeta$,
$\xi =\xi_0+\xi_1(\tilde{x},\tilde{y})$,
$\zeta =\zeta_0+\zeta_1(\tilde{x},\tilde{y})$, where $\xi_1\ll\xi_0$,
$\zeta_1 \ll \zeta_0$.
The richest distinguished limit corresponds to having
$\gamma = \frac{4}{3}\delta + O(\delta^{3/2})$,
$\xi_1, \, \zeta_1 = O(\sqrt{\delta})$, $\tilde{x},\, \tilde{y} =
O(\delta^{-1/2})$, and $\tilde{z}=O(\delta^{-1})$.
Inserting this Ansatz into Eqs.\ (\ref{half}), the terms
$\partial_{\tilde{z} \tilde{z}} \tilde{b}$ and
$\partial_{\tilde{z}\tilde{z}} \tilde{c}$ are $O(\delta^{5/2})$
and can be ignored when compared with the others, which are
$O(\delta^{3/2})$.
Inserting the result into (\ref{screen-full}), we find

\vspace*{0.2cm}

\begin{mathletters}\begin{eqnarray} \label{adiabat-a}
\partial_{\tilde{x}\tilde{x}} \xi_1  + \partial_{\tilde{y}\tilde{y}}
\xi_1 = \case{\delta}{\sqrt{3}}\;\xi_1 &-& \case{32}{27}\;\xi_1^3
\nonumber\\
&+& \case{\sqrt{3}}{4}\; (\case{3}{4}\gamma - \delta) +
O(\delta^{5/2}), \\ &&\zeta_1 = \pm\xi_1/\sqrt{3} + O(\delta),
\label{adiabat-b} \end{eqnarray} \end{mathletters}
Notice that our Ansatz corresponds to weakly nonlinear perturbations
of uniform solutions varying on a large spatial scale
$\tilde{x}$=$\sqrt{2} kx = O(\delta^{-1/2})\gg 1$.
The typical transverse length over which our solutions vary
is thus much larger than the wavelength $1/k$.

With the substitutions: $\xi_1 = \frac{3}{4} 3^{1/4}\delta^{1/2} u$,
$\tilde{x} = 3^{1/4} \delta^{-1/2} X$,
$\tilde{y}$=$3^{1/4}\delta^{-1/2} Y$,
Eq.\ (\ref{adiabat-a}) can be written in the simpler form:
$\partial_{XX} u + \partial_{YY} u = u - 2 u^3 + \mu + O(\delta)$,
$\mu = 3^{-1/4}\delta^{-3/2} (\frac{3}{4}\gamma - \delta) = O(1)$.
We report here only the results (for $y$-independent solutions and for
the axisymmetric case) corresponding to the most symmetrical situation
$\gamma=\frac{4}{3}\delta$ ($\mu$=0), where explicit formulas can be
easily obtained.
The results for the general non-symmetric case will be published
elsewhere.
 
\paragraph{Two-dimensional solutions depending on one transversal
coordinate.}
If $u$=$u(X)$ (two-dimensional solutions of the full problem depending
on only one transversal coordinate), the parameter-free equation
$\partial_{XX} u=u-2u^3$ can be integrated once yielding the result
$(\partial_X u)^2=u^2-u^4+C$.
This equation admits nonuniform solutions satisfying the condition of
zero flux as $X\to\pm\infty$ only if $C$=0. In this case we obtain the
solutions $u=\eta\;{\rm sech}(X-X_0)$, with $\eta$=1 for the soliton
and $\eta$=$-1$ for the antisoliton. Next we choose $X_0$=0.
Finally, using relation (\ref{adiabat-b}) our solution for the
transmitted and reflected amplitudes on the screen will be

\begin{mathletters} \label{soliton} \begin{eqnarray}
\tilde{c}(\tilde{x}) &=& \frac{\sqrt{3}}{2}\; e^{\mp i \pi/6}
[1+\sqrt{3}\eta\lambda\;{\rm sech}(\lambda\tilde{x})\;
e^{\pm i \pi/3}], \\ \tilde{b}(\tilde{x}) &=& -\frac{1}{2}\;
e^{\pm i \pi/3} [1-3\sqrt{3}\eta\lambda\;
{\rm sech}(\lambda\tilde{x})\; e^{\mp i \pi/6}],
\end{eqnarray} \end{mathletters}

\noindent where $\lambda$=$3^{-1/4}\delta^{1/2}$ and the upper and
lower signs refer to the cases $\cal A$ and $\cal R$, respectively.
On the screen $z$=0 the difference between the cases $\cal A$ and
$\cal R$ lies only in the phase of the wave function and not in the
intensities:  $|\tilde{c}(\tilde{x})|^2=
\frac{3}{4}[1+\sqrt{3}\eta\lambda\;{\rm sech}(\lambda\tilde{x})]$,
$|\tilde{b}(\tilde{x})|^2= \frac{1}{4}[1-3\sqrt{3}\eta\lambda\;
{\rm sech}(\lambda\tilde{x})]$ (the small terms $\sim\lambda^2$ are
dropped consistently with our scaling). The different phase factors
give rise to drastic differences in the wave function outside the
screen, as it will be shown below. We have also checked that the
solutions are linearly stable when time evolution is considered
subject to the boundary conditions discussed earlier.

The amplitudes of the transmitted and reflected waves outside
the screen can be found from (\ref{half}) using as the boundary
conditions their values at $z$=0 and ignoring the small terms
$\partial_{\tilde{z}\tilde{z}}\tilde{c}$ and
$\partial_{\tilde{z}\tilde{z}}\tilde{b}$:

\begin{equation} \label{diff-ampl}
\tilde{c}(\tilde{x},\tilde{z})=\frac{1}{\sqrt{4\pi i\tilde{z}}}
\int_{-\infty}^{\infty} \tilde{c}(\tilde{x}',0)\;
e^{i (\tilde{x}-\tilde{x}')^2/(4\tilde{z})}\; d \tilde{x}',
\end{equation}
and the expression for $\tilde{b}(\tilde{x},\tilde{z})$ is the same
once $\tilde{z}$ is replaced by $-\tilde{z}$.

In our 2D problem the intensities of the reflected and transmitted
waves are nonuniform in space in contrast to the 1D problem, where
they are constant. Denoting
$|\tilde{c}(\tilde{x},\tilde{z})|^2=\Psi_t^0
\{1+\Psi_t(\tilde{x},\tilde{z})\}$,
$|\tilde{b}(\tilde{x},\tilde{z})|^2=\Psi_r^0
\{1+\Psi_r(\tilde{x},\tilde{z})\}$, and using for
$\tilde{c}(\tilde{x},0)$, $\tilde{b}(\tilde{x},0)$ the soliton-type
solutions (\ref{soliton}), we obtain $\Psi_t^0$=3/4, $\Psi_r^0$=1/4.
The nonuniform parts of the intensities are given by
\begin{eqnarray} \label{diff-int}
\Psi_{r,t}&&(\tilde{x},\tilde{z}) = \nonumber \\
&&\eta\frac{\sigma_{r,t}\lambda}{\sqrt{\pi |\tilde{z}| }}
\int_{-\infty}^{\infty}\cos[\frac{(\tilde{x}-\tilde{x}')^2}
{4 |\tilde{z}|} +\phi_{r,t}^{\pm}] \;{\rm sech}(\lambda\tilde{x}')
d \tilde{x}'
\end{eqnarray}
where $\sigma_{r}$=$-3$, $\sigma_{t}$=$\sqrt{3}$, and the terms
$\sim\lambda^2$ were dropped.
The phase $\phi$ in the argument of cosine is different for
the attractive and repulsive nonlinearities:  (case $\cal A$):
$\phi_t^{+}=\frac{1}{12}\pi$, $\phi_r^{+}=-\frac{5}{12}\pi$;
(case $\cal R$): $\phi_t^{-}=-\frac{7}{12}\pi$,
$\phi_r^{-}=-\frac{1}{12}\pi$.

\begin{figure}
\vspace*{-1cm}
\setlength{\epsfxsize}{12cm}
\centerline{\mbox{\epsffile{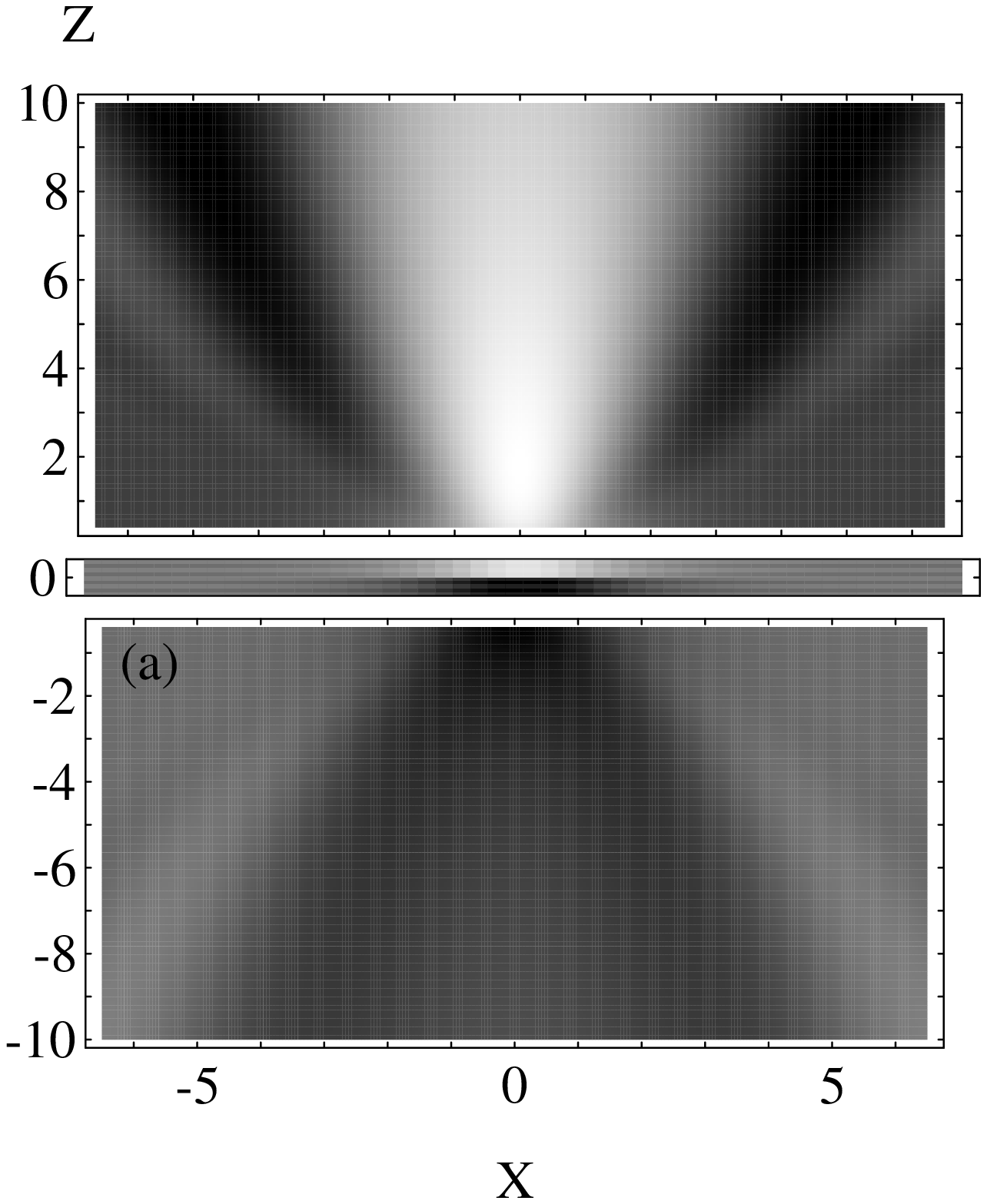}}}
\vspace*{-6.5cm}
\setlength{\epsfxsize}{12cm}
\centerline{\mbox{\epsffile{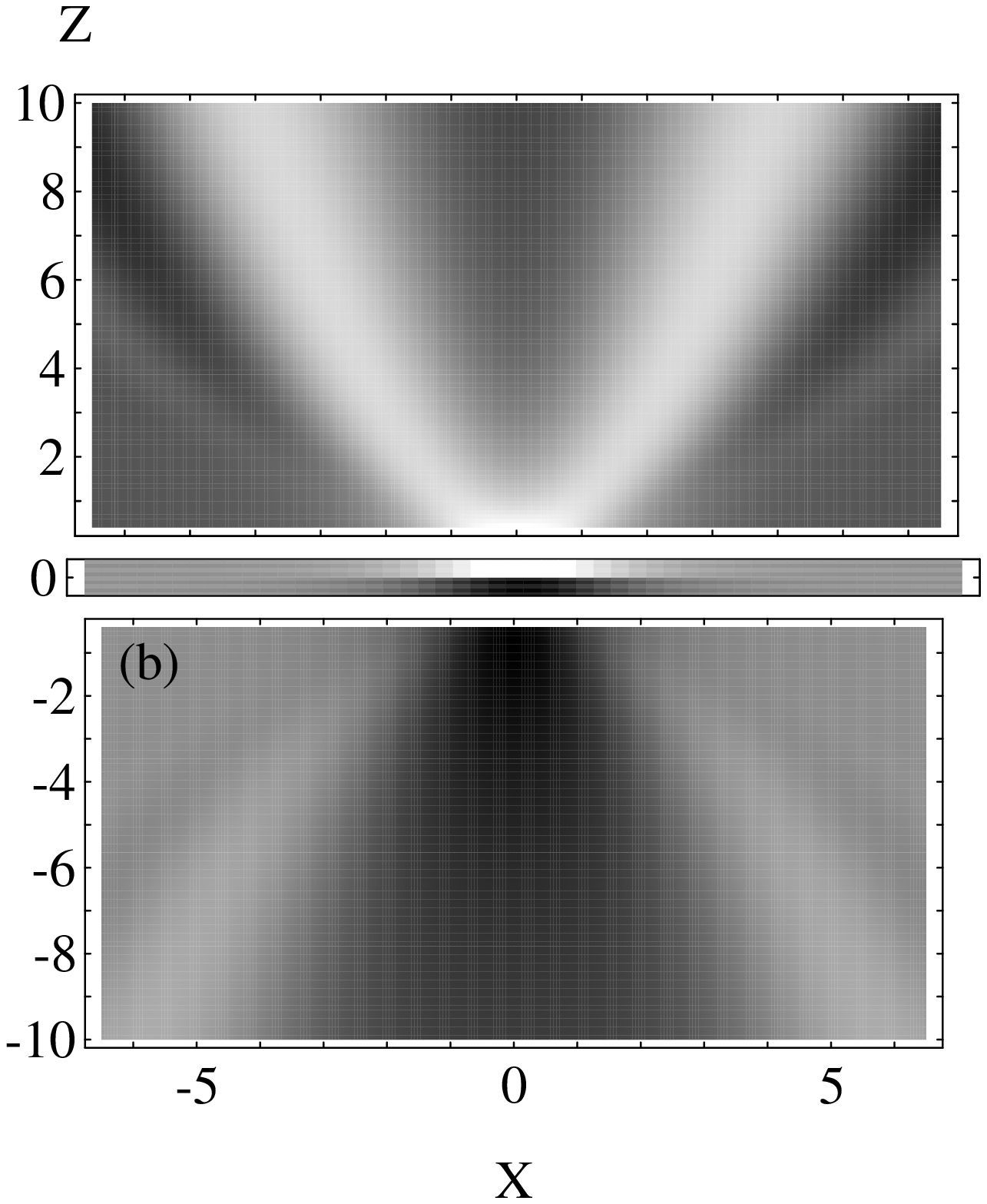}}}
\vspace*{-4cm}
\caption{
Density plots for the wave function intensities created by scattering
off the self-induced nonuniform pattern on the screen at $Z$=0 with
attractive (a) or repulsive (b) nonlinearities.
White (black) color corresponds to the maximum (minimum) of the
intensity for the soliton solution on the screen ($\eta =1$), and vice
versa for the antisoliton solution ($\eta$=$-1$).
}\label{diffraction}\end{figure}

Spatial distributions of the wave intensities are obtained by
numerical integration of Eq.\ (\ref{diff-int}) and presented
in Fig.\ \ref{diffraction} in terms of the scaled coordinates
$X=\lambda\tilde{x}$, $Z=4\lambda^2\tilde{z}$. The off-screen wave
intensities are shown starting from certain nonzero values of $Z$.
The wave intensity on the screen is shown as a thin strip in the
middle of Figs.\ \ref{diffraction}(a),(b).
The results can be interpreted as follows.

The uniform incident flow spontaneously produces a nonuniform
soliton-type pattern on the screen and then is diffracted by it
due to the nonlinear feedback in the equations.
In particular, for the soliton solution ($\eta$=1) we observe
{\em local self-brightening} of the transmitted wave with simultaneous
{\em local suppression} of the reflected wave
(Fig.\ \ref{diffraction}).
The diffraction pattern is crucially determined by the value of the
phase factor $\phi^{\pm}$.
For the case $\cal A$ the transmitted wave is focused into a ``beam''
of higher intensity with a maximum outside the screen at $Z\approx
1.7$ [Fig.\ \ref{diffraction}(a)], whereas for the case $\cal R$ it is
defocused and it ``splits'' into two ``beams''
[Fig.\ \ref{diffraction}(b)].
Additional support for the importance of the phase factor is provided
by the asymptotic behavior of the integral (\ref{diff-int})
in the remote zone $Z\gg 1$, $2X\ll Z$

\begin{equation} \label{diff-ass}
\Psi_{r,t}(\tilde{x},\tilde{z}) \approx
\eta\;\sigma_{r,t}\sqrt{\frac{\pi}{|\tilde{z}|}}
\cos(\frac{\tilde{x}^2}{4 |\tilde{z}|}+\phi_{r,t}^{\pm}).
\end{equation}
In the transverse direction the local maxima (minima) of the
intensities are determined by the condition $\frac{\tilde{x}^2}
{4 |\tilde{z}|}+\phi_{r,t}^{\pm}=\pi m$, $m=0,\pm 1,\dots$.
For instance $\phi_t^{+} > 0$ for case $\cal A$, and the cosine in
(\ref{diff-ass}) reaches its maximum at $\tilde{x}$=0 providing a
transmitted ``beam'' along the axes.
$\phi_t^{-} < 0$ for case $\cal R$, and the cosine is largest
on the parabola $\tilde{z}=\frac{3}{7\pi}\tilde{x}^2$ yielding
two transmitted ``beams'' as in Fig.\ \ref{diffraction}(b).
The behavior of the reflected wave is also nontrivial:
for case ${\cal A}$ the reflected pattern contains ``split traces'':
the suppressed reflection forms the parabola
$\tilde{z}=\frac{3}{5\pi}\tilde{x}^2$ [Fig.\ \ref{diffraction}(a)],
whereas for case ${\cal R}$ the reflection is suppressed within a
single trace [Fig.\ \ref{diffraction}(b)].
It should be noted, however, that for the antisoliton solution
($\eta$=$-1$) the maxima and minima are interchanged (with respect to
the soliton solution), the ``beams'' become the suppressed traces and
vice versa.
Which type of solution (self-brightening or self-darkening of the
transmission) will be realized in practice depends on additional
conditions: type of the imperfections pinning the soliton, boundary
conditions, past history, and so on.

\begin{figure}
\vspace*{0.5cm}
\setlength{\epsfxsize}{8cm}
\centerline{\mbox{\epsffile{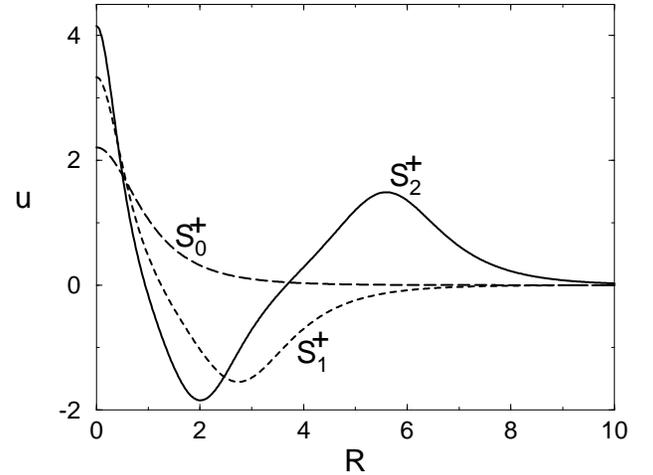}}}
\caption{
Axisymmetric solutions $S_n^{+}$ for the transmitted amplitude on the
screen when n=1,2,3. Recall that $S_n^{-}$=$-S_n^{+}$.
}\label{axisym}\end{figure}

\paragraph{Two-dimensional axisymmetric solutions.}
For this geometry, $u$=$u(R)$, $R^2=X^2+Y^2$,
and the parameter-free equation becomes
$\partial_{RR}u+\frac{1}{R}\partial_R u=u-2u^3$.
There are infinitely many solutions which satisfy our boundary
conditions $u(\infty) =0$ and $u_R(0)$=$u_R(\infty)$=0.
We denote them by $S_n^{+}$ for the soliton case [$u(0)>0$] and
$S_n^{-}$ for the antisoliton case [$u(0)<0$], where the subscript
$n=0,1,\dots$ is the number of zeros of $S_n^{\pm}$ as shown
in Fig.\ \ref{axisym}. Then $S_n^{-}$=$-S_n^{+}$.

In conclusion, spontaneous formation of spatial transverse patterns,
which are quantum-mechanically coherent is expected to occur in
semiconductor heterostructures with a thin nonlinear layer.
Self-diffraction of the electron wave on the transverse patterns gives
rise to interesting phenomena like self-brightening or darkening of
the transmitted wave, beam splitting, etc.

This work has been supported by the DGICYT grants PB92-0248 and
PB94-0375, and by the EU Human Capital and Mobility Programme contract
ERBCHRXCT930413. O.M.B. acknowledges support from the Ministerio de
Educaci\'on y Ciencia of Spain.

\end{document}